\documentclass[aps,pra,nopacs,oneside,floatfix,a4,superscriptaddress,twocolumn,nofootinbib]{revtex4-1}
\usepackage{bbm}
\usepackage{amsmath, amssymb,amsthm}
\usepackage{color}
\usepackage{graphicx}
\usepackage{times}
\usepackage{tikz}
\usepackage{mathptmx}
\theoremstyle{plain}

\theoremstyle{definition}

\usepackage[space]{grffile}
\usepackage{latexsym}
\usepackage{textcomp}
\usepackage{multirow,booktabs}
\usepackage{amsfonts,amsmath,amssymb}
\usepackage{url}
\usepackage{hyperref}
\hypersetup{colorlinks=false,pdfborder={0 0 0}}
\usepackage[utf8]{inputenc}
\usepackage[english]{babel}

\usepackage[normalem]{ulem}
\usepackage{multirow}
\usepackage{times}

\usepackage{amsfonts}
\usepackage{latexsym}
\usepackage{amsmath}
\usepackage{amssymb}
\usepackage{amsfonts}
\usepackage{amsthm}
\usepackage{mathrsfs}
\usepackage{natbib}
\usepackage{color,verbatim,graphics}
\usepackage{psfrag}
\DeclareMathAlphabet{\mathrsfs}{U}{rsfs}{m}{n}
\DeclareMathAlphabet{\mathpzc}{OT1}{pzc}{m}{it}
\DeclareMathAlphabet{\matheus}{U}{eus}{m}{n}
\DeclareMathAlphabet{\mathbbold}{U}{bbold}{m}{n}

\newcommand{\ba}{\begin{eqnarray}}
\newcommand{\be}{\begin{equation}}
\newcommand{\ee}{\end{equation}}
\newcommand{\beq}{\begin{equation}}
\newcommand{\eeq}{  \end{equation}}
\newcommand{\bea}{\begin{eqnarray}}
\newcommand{\eea}{  \end{eqnarray}}

\newcommand{\ea}{\end{eqnarray}}
\newcommand{\ban}{\begin{eqnarray*}}
\newcommand{\ean}{\end{eqnarray*}}
\newcommand{\Tr}{\operatorname{Tr}}
\newcommand{\tr}{\operatorname{Tr}}

\newcommand{\ket}[1]{\left|#1\right\rangle}
\newcommand{\bra}[1]{\langle#1|}

\newcommand{\ie}{{\it{i.e.}~}}

\newcommand{\tp}{\otimes}

\definecolor{matty}{rgb}{1,0,0}

\begin{document}

\title{Measurement-device-independent entanglement and randomness estimation in quantum networks}

\author{Ivan \v{S}upi\'{c}}
\affiliation{ICFO-Institut de Ciencies Fotoniques, The Barcelona Institute of Science and Technology, 08860 Castelldefels (Barcelona), Spain}
\email{ivan.supic@icfo.eu}

\author{Paul Skrzypczyk}
\affiliation{H. H. Wills Physics Laboratory, University of Bristol, Tyndall Avenue, Bristol, BS8 1TL, United Kingdom}

\author{Daniel Cavalcanti}
\affiliation{ICFO-Institut de Ciencies Fotoniques, The Barcelona Institute of Science and Technology, 08860 Castelldefels (Barcelona), Spain}

\date{\today}

\begin{abstract}
Detection of entanglement in quantum networks consisting of many parties is one of the important steps towards building quantum communication and computation networks. We consider a scenario where the measurement devices used for this certification are uncharacterised. In this case, it is well known that by using quantum states as inputs for the measurement devices it is possible to detect any entangled state (a situation known as measurement device-independent entanglement witnessing). Here we go beyond entanglement detection and provide methods to estimate the amount of entanglement in a quantum network. We also consider the task of randomness certification and show that randomness can be certified in a variety of cases, including single-partite experiments or setups using only separable states.
\end{abstract}

\maketitle

\section{Introduction}

Quantum entanglement is one of the most intriguing aspects of quantum theory \cite{QE}. Implications of its existence on the foundations of quantum theory were already emphasized in the seminal work of Einstein, Podolsky, and Rosen \cite{EPR}. In the last few decades entanglement has been at the core of the power of quantum theory for different information processing tasks \cite{QE}. Moreover, it allows quantum systems to display nonlocal correlations, which can be certified by the violation of Bell inequalities \cite{Bell, NLreview}. Thus, whenever a Bell inequality violation is observed one certifies that the underlying quantum system is entangled. Importantly, this certification is done without the need to know which measurements are being performed (in contrast to the case of entanglement witnesses \cite{QE}). We can thus say that the observation of nonlocal correlations certifies entanglement in a device-independent way \cite{NLreview}. \\

The relation between Bell nonlocality and entanglement is one of the open questions in the foundations of quantum theory. While nonlocal correlations can only be obtained by performing local measurements on an entangled state, not every entangled state can lead to nonlocality. As it was first shown by Werner \cite{Werner} the probability distributions obtained by local measurements applied to certain entangled quantum states can be simulated by a purely classical model called a local-hidden-variable model. Werner's result was latter generalised in many ways \cite{Barrett,LHV1,LHV2,LHV3,LHV4,LHV5,LHV6, LHV7,LHV8,LHV9,LHV10,LHV11}.\\

Since the standard Bell scenario exposes a gap between nonlocality and entanglement, one may ask if it is possible to come up with some alternative scenario in which every entangled state will exhibit nonlocal correlations. One possibility is to make use of many copies of the shared entangled state. It has been proven that some entangled states which are not nonlocal in the standard Bell scenario can violate a Bell inequality in this scenario \cite{Palazuelos,ManyCopiesUs}, a phenomenon known as super-activation. Whether every entangled state can be super-activated in this way is again an open problem. Another possibility is to consider a quantum network consisting of many copies of the same state shared among many parties \cite{Sen,Dani1,Dani2}. Similarly to superactivation, some states that are local in the standard Bell scenario become nonlocal in the network scenario. However, the general relation between entanglement and nonlocality is also unknown in this case. Another alternative Bell scenario, historically preceding the others, was suggested in \cite{Popescu} (see also \cite{LHV11}) and is known as the hidden nonlocality scenario. In it the parties are allowed to perform pre-processing on their shared state before applying their measurements. Even though some local entangled states can become nonlocal in this new scenario it has recently been shown that there are entangled states which can never exhibit hidden nonlocality, i.e. they remain local after arbitrary pre-processing \cite{Flavien}. Finally, it is possible to combine all the nonstandard Bell scenarios, but still there is no conclusive statement about the relation between entanglement and nonlocality in what would be called multy-copy with pre-processing scenario.

There is however one modification of the standard Bell scenario that can reveal nonlocal correlations from every entangled state. It consists of using measurement devices that receive quantum systems as inputs \cite{Buscemi} (See also \cite{Branciard, Rosset, Ecavalcanti,Hall16} for further developments and \cite{Nawareg,Xu,Verbanis} for experimental demonstrations). The observation of nonlocal correlations in this scenario can be seen as an entanglement test with uncharacterised measurements, which motivated the name of \emph{measurement-device-independent nonlocality}. 

The aim of this paper is to shed new light on some aspects of this new scenario with quantum inputs, explore its power for entanglement detection and quantification in quantum networks, and finally to study randomness certification in this scenario. 

\section{Measurement-device-independent entanglement certification}

We start by reviewing some of the main results on entanglement certification in Bell scenario with quantum inputs, and by proposing some improvements on this task. We consider two separated parties, Alice and Bob, sharing a bipartite system in an unknown state $\rho^{AB} \in \mathcal{L}(\mathcal{H}_A \otimes \mathcal{H}_B)$. They want to certify if their system is entangled, but do not know how their measurement devices work. Buscemi proposed a solution to this problem \cite{Buscemi}: at each round of the experiment Alice and Bob encode their measurement choices in quantum states $\psi_x \in \mathcal{L}(\mathcal{H}_{A_0})$ and $\psi_y \in \mathcal{L}(\mathcal{H}_{B_0})$ respectively, which they use as inputs for their measurement devices. After receiving the quantum inputs the measurement devices provide classical outputs $a$ and $b$ according to the probability distributions %
\begin{multline}\label{eq:qinlpd}
p(a,b\vert \psi_x,\psi_y) =
\textrm{Tr}[(M_{a}^{A_0A} \otimes M_{b}^{BB_0})(\psi_x^{A_0}\otimes \rho^{AB}\otimes \psi_y^{B_0})],
\end{multline}
where $M_{a}^{A_0A}$ and $M_{b}^{BB_0}$ are the measurement operators applied by the measurement devices on the corresponding input systems and their shares of state $\rho^{AB}$. Buscemi proved that if $\{\psi_{x/y}\}_{x/y}$  correspond to tomographically complete sets of input states \footnote{By a tomographically complete set of input states we mean that the set is sufficient to perform quantum process tomography} in $\mathcal{H}_{A/B}$, and each box performs a Bell state measurement, then every entangled state $\rho^{AB}$ produces a set of probability distributions $\{p(a,b\vert \psi_x,\psi_y)\}_{a,b,x,y}$ that cannot be reproduced with any separable state.\\

Building on Buscemi's result, the authors of \cite{Branciard} provided a method of constructing Bell tests with quantum inputs from every entanglement witness, and named them measurement-device-independent entanglement witnesses (MDIEW). However, when it comes to practical entanglement detection, MDIEW are useful only when one has a good guess on which entangled state should be detected, and can thus start from an entanglement witness which is able to detect its entanglement. This problem was first addressed in \cite{Mallick}, where it was shown that given multiple copies of the state then a universal witness can be found. More recently, in \cite{Verbanis} a solution at the single copy level was given, by showing that the quantum inputs scenario can be cast as a semidefinite programming (SDP) optimization problem \cite{SDP}, readily solvable with available software. We present here the said SDP in a slightly different form.

The starting point is the fact that the joint outcome probability distribution can be written in the following way
\be\label{eq:effectiveProb}
p(a,b\vert \psi_x,\psi_y) = \textrm{Tr}[\tilde{M}_{a,b}^{A_0B_0}(\psi_x^{A_0}\otimes \psi_y^{B_0})],
\ee
where $\tilde{M}_{a,b}$ is an effective POVM operator defined by
\be\label{eq:effectivePOVM}
\tilde{M}_{a,b}^{A_0B_0} = \textrm{Tr}_{AB}[(M_{a}^{A_0A} \otimes M_{b}^{BB_0})(\mathbb{I}^{A_0}\otimes \rho^{AB}\otimes \mathbb{I}^{B_0})].
\ee
By construction, the effective POVM $\tilde{M}_{a,b}^{A_0B_0}$ satisfies a number of conditions, which can be thought of as playing the role of `no-signalling' conditions. In particular, 
\begin{equation}
	\begin{split}
		\sum_a \tilde{M}_{a,b}^{A_0B_0} = \mathbb{I}^{A_0} \otimes \tilde{M}_b^{B_0} \\
		\sum_b \tilde{M}_{a,b}^{A_0B_0} = \tilde{M}_a^{A_0} \otimes \mathbb{I}^{B_0} 
	\end{split}
\end{equation}
where $\tilde{M}_b^{B_0} \equiv \tr_B[M_b^{BB_0}(\rho^B\otimes \mathbb{I}^{B_0})] \geq 0$ and $\tilde{M}_a^{A_0} \equiv \tr_A[M_a^{A_0A}(\mathbb{I}^{A_0}\otimes \rho^A)] \geq 0$ are effective local POVMs for Alice and Bob (i.e. such that $\sum_a \tilde{M}_a^{A_0} = \mathbb{I}^{A_0}$ and $\sum_b \tilde{M}_b^{B_0} = \mathbb{I}^{B_0}$). We will write  $\{\tilde{M}_{a,b}^{A_0B_0}\}_{a,b} \in \mathcal{M}$ to denote the fact that the effective POVM satisfies these conditions, i.e.
\begin{multline}\label{e:set M}
\mathcal{M} = \Big\{\{\tilde{M}_{a,b}^{A_0B_0}\}_{a,b}|\tilde{M}_{a,b}^{A_0B_0} \geq 0, \sum_a \tilde{M}_{a,b}^{A_0B_0} = \mathbb{I}^{A_0} \otimes \tilde{M}_b^{B_0}, \\
\sum_b \tilde{M}_{a,b}^{A_0B_0} = \tilde{M}_a^{A_0} \otimes \mathbb{I}^{B_0}, \sum_a \tilde{M}_a^{A_0} = \mathbb{I}^{A_0},\sum_b \tilde{M}_b^{B_0} = \mathbb{I}^{B_0}\Big\}
\end{multline}

Now, if the shared state $\rho^{AB}$  is separable, \ie  $\rho^{AB}=\sum_{\lambda}p_{\lambda}\rho_{\lambda}^A\otimes \rho^B_{\lambda}$, then \eqref{eq:effectivePOVM} becomes 
\begin{align}\label{eq:effectivePOVMsep}
\tilde{M}_{a,b}^{A_0B_0} &= \sum_{\lambda}p_{\lambda}\textrm{Tr}_{A}[M_{a}^{A_0A} (\mathbb{I}^{A_0}\otimes \rho_{\lambda}^A)]\otimes \textrm{Tr}_{B} [M_{b}^{BB_0}(\rho^B_{\lambda}\otimes \mathbb{I}^{B_0})]\nonumber\\
&= \sum_{\lambda}p_{\lambda} \tilde{M}_{a|\lambda}^{A_0}\otimes \tilde{M}_{b|\lambda}^{B_0}.
\end{align}
where $\tilde{M}_{a|\lambda}^{A_0}\equiv \textrm{Tr}_{A}[M_{a}^{A_0A} (\mathbb{I}^{A_0}\otimes \rho_{\lambda}^A)]$ and $\tilde{M}_{b|\lambda}^{B_0}\equiv \textrm{Tr}_{B} [M_{b}^{BB_0}(\rho^B_{\lambda}\otimes \mathbb{I}^{B_0})]$ are effective local POVMs for Alice and Bob. Consequently, the fact that $\rho^{AB}$ is separable, implies that the operator $\tilde{M}_{a,b}^{A_0B_0}$ is a separable operator for all $a$ and $b$.

Alice and Bob can thus check the separability of $\rho^{AB}$ by solving the following feasibility problem: 
\begin{align}
\textrm{given}& \quad \{p(a,b\vert \psi_x,\psi_y)\}_{a,b,x,y} \nonumber \\
\textrm{find}& \quad \{\tilde{M}^{A_0B_0}_{a,b}\}_{a,b} \label{eq:sdp} \\
\textrm{s.t.}&\quad p(a,b\vert \psi_x,\psi_y) = \textrm{Tr}[\tilde{M}_{a,b}^{A_0B_0}(\psi_x^{A_0}\otimes \psi_y^{B_0})] \quad \forall a,b,x,y, \nonumber \\
&\quad \{\tilde{M}_{a,b}^{A_0B_0}\}_{a,b} \in \mathcal{S} \nonumber
\end{align}
where $\mathcal{S}$ denotes the subset of $\mathcal{M}$ which are separable operators, \ie 
\begin{multline}\label{e:set S}
\mathcal{S} = \Big\{\{\tilde{M}_{a,b}^{A_0B_0}\}_{a,b}|\{\tilde{M}_{a,b}^{A_0B_0}\}_{a,b} \in \mathcal{M}, \\ \tilde{M}_{a,b}^{A_0B_0}= \sum_{\lambda} \tau_{a|\lambda}\otimes \chi_{b|\lambda},
\tau_{a|\lambda} \geq 0, \chi_{b|\lambda} \geq 0\Big\}
\end{multline}

This problem is in principle  hard to solve, due to the lack of an efficient characterization of the set of separable operators. However one can relax the constraint of separability, and impose instead that each operator $\tilde{M}_{a,b}^{A_0B_0}$ is positive under partial transpose (PPT). In the feasibility problem above this amounts to replacing the condition $\tilde{M}_{a,b}^{A_0B_0}= \sum_{\lambda} \tau_{a|\lambda}\otimes \chi_{b|\lambda},
\tau_{a|\lambda} \geq 0, \chi_{b|\lambda} \geq 0$ by $(\tilde{M}_{a,b}^{A_0B_0})^{T_{A_0}}\geq0~\forall~a,b$. With this replacement, the problem becomes a feasibility SDP optimisation problem, which can then be solved efficiently.\\ 

Notice however that this relaxation is not able to detect PPT entangled states. A second, more stringent, relaxation of the set of separable operators is the set of the operators having a $k$-symmetric extension \cite{Doherty}. Imposing that the operators $\tilde{M}_{a,b}^{A_0B_0}$ have a $k$ symmetric extension amounts to demanding that there exist a $(k+1)$-partite operator $\tilde{N}_{a,b}^{A_0B_0...B_{k-1}}\geq0$ such that $\tilde{N}_{a,b}^{A_0B_i}=\tilde{M}_{a,b}^{A_0B_0}~\forall ~i$. For every fixed $k$, the above feasibility optimisation problem with this replacement is again a SDP feasibility problem,  which now can also detect PPT entangled states \cite{Doherty1}. Finally,  we note that by increasing the order $k$ of the extension, we obtain stronger SDP tests that converge to the separability test above in the limit of $k\rightarrow\infty$.\\ 

Finally, it is an important fact that once the sets of quantum inputs used are tomographically complete, then the probabilities $p(a,b\vert \psi_x,\psi_y)$ allow for an exact reconstruction of the effective POVM elements $\tilde{M}_{a,b}^{A_0B_0}$, using quantum process tomography. In this special case, \textit{any} available entanglement criterion, not just those that can be checked via SDP, can be used to determine if it is a separable operator or not,  leading directly to a conclusion of whether or not the shared state was entangled. 

\section{Measurement-device-independent entanglement estimation}

One step beyond certifying the presence of entanglement in a system is to estimate how much entanglement it contains. There are many different entanglement measures, but they are in general not easy to compute even if one knows the full state of the system (for more details on entanglement measures see \cite{Plenio}). In fact the problem of deciding if a given quantum state is entangled is NP-hard, which implies hardness of computing entanglement measures \cite{NP-hard}. Quantification of entanglement in a measurement-device-independent scenario was the subject of \cite{Shahandeh} where the authors define the best possible pay-off in a semiquantum games that a state can achieve as an entanglement measure. In what follows we show how to place measurement-device-independent bounds on two well-known entanglement quantifiers, the robustness \cite{Vidal} and the negativity \cite{Negativity}. 

\subsection{MDI lower bound on the robustness of entanglement}

A physically well motivated way of quantifying entanglement is through its robustness to noise \cite{Vidal}, defined as the amount of noise one can add to an entangled state before it becomes separable. In mathematical terms, the generalised robustness $r_g$ of a state $\rho$ is given by
\begin{align}\label{entrob}
r_\mathit{S}(\rho^{AB}) = \min_{r,\sigma^{AB}} &\quad r \\ \nonumber
\textrm{s.t.}&\quad \frac{\rho^{AB} + r\sigma^{AB}}{1+r} \in \mathit{SEP}, \nonumber \\
&\quad\sigma^{AB} \in S
\end{align}
where $\mathit{S}$ is a subset of quantum states, which defines the type of robustness, and $\mathit{SEP}$ denotes the set of separable states. Typical choices for $\mathit{S}$ include the set of all quantum states (generalised robustness), the set of separable states (robustness) or the maximally mixed state (random robustness).  

In a similar way we define the robustness of MDI-nonlocality $\tilde{r}^{MDI}_\mathit{S}$ as the minimum amount of noise that has to be added to the set of probability distributions $\{p(a,b\vert \psi_x,\psi_y)\}_{a,b,x,y}$ before it can be reproduced by a separable state, where the noise comes from the set $\mathit{S}$. Formally, the MDI-nonlocality robustness is the solution of the following optimization problem
\begin{align}\label{mdirob}
&\tilde{r}^{MDI}_\mathit{S}[p(a,b\vert \psi_x,\psi_y)] = \min_{\{\tilde{M}_{a,b}^{A_0B_0}\}_{a,b},\{\tilde{N}_{a,b}^{A_0B_0}\}_{a,b}}\quad r \\ 
& \textrm{s.t.} \quad \frac{p(a,b\vert \psi_x,\psi_y) + r\pi(a,b\vert \psi_x,\psi_y)}{1+r} = \textrm{Tr}[\tilde{M}_{a,b}^{A_0B_0}(\psi_x^{A_0}\otimes \psi_y^{B_0})], \nonumber \\
&\qquad\pi(a,b\vert \psi_x,\psi_y) = \textrm{Tr}[\tilde{N}_{a,b}^{A_0B_0}(\psi_x^{A_0}\otimes \psi_y^{B_0})] \quad \forall a,b,x,y, \nonumber \\
&\qquad\{\tilde{M}_{a,b}^{A_0B_0}\}_{a,b} \in \mathcal{S}, \qquad\{\tilde{N}_{a,b}^{A_0B_0}\}_{a,b} \in \mathcal{M}_\mathit{S}. \nonumber
\end{align}
where $\mathcal{M}_\mathit{S}$ is the set of effective POVMs associated to the noise $\mathit{S}$. For example, for the generalised robustness, when the set $\mathit{S}$ corresponds to all quantum states, then $\mathcal{M}_S = \mathcal{M}$. Similarly, for the robustness, when the set $\mathit{S}$ corresponds to all separable states, then $\mathcal{M}_\mathit{S} = \mathcal{S}$. Finally, for the random robustness, when $\mathit{S} = \{\mathbb{I}^{AB}/d_Ad_B\}$, then $\mathcal{M}_\mathit{S} = \Big\{\{\tilde{M}_{a,b}^{A_0B_0}\}_{a,b}|\{\tilde{M}_{a,b}^{A_0B_0}\}_{a,b} \in \mathcal{M}, \tilde{M}_{a,b}^{A_0B_0} = \tilde{M}_a^{A_0} \otimes \tilde{M}_b^{B_0}\Big\}$. 

We now show that $\tilde{r}^{MDI}_\mathit{S}$ is a lower bound to the robustness of entanglement $r_\mathit{S}$ of the underlining state being measured. To see this consider that the robustness of the state $\rho^{AB}$ is given by $r_\mathit{S}^*$. This means that there exist a state ${\sigma^*}^{AB} \in \mathit{S}$ for which the state $(\rho + r^*_\mathit{S}\sigma^*)/(1+r^*_\mathit{S})$ is separable. Thus for any POVMs $\{M_a^{A_0A}\}_a$ and $\{M_b^{BB_0}\}_b$ satisfying \eqref{eq:qinlpd}, 
\begin{equation}\label{e:rg M}
\tilde{M}_{a,b}^{A_0B_0} = \textrm{Tr}_{AB}\left[\left(M_{a}^{A_0A} \otimes M_{b}^{BB_0}\right)\left(\mathbb{I}^{A_0}\otimes \frac{\rho^{AB}+r^*_\mathit{S}{\sigma^*}^{AB}}{1+r^*_\mathit{S}}\otimes \mathbb{I}^{B_0}\right)\right],
\end{equation}
and 
\begin{equation}\label{e:rg N}
\tilde{N}_{a,b}^{A_0B_0} = \textrm{Tr}_{AB}\left[\left(M_{a}^{A_0A} \otimes M_{b}^{BB_0}\right)\left(\mathbb{I}^{A_0}\otimes {\sigma^*}^{AB}\otimes \mathbb{I}^{B_0}\right)\right],
\end{equation}
are feasible for the problem \eqref{mdirob} (i.e. satisfy all the constraints) and achieve the value $r = r_\mathit{S}^*$, as can be verified by direct substitution. Since $\{\tilde{M}_{a,b}^{A_0B_0}\}_{a,b}$ and $\{\tilde{N}_{a,b}^{A_0B_0}\}_{a,b}$ given by \eqref{e:rg M} and \eqref{e:rg N} do not necessarily provide an optimal solution to the problem \eqref{mdirob}, then 
\be
\tilde{r}^{MDI}_\mathit{S}[p(a,b\vert \psi_x,\psi_y)] \leq r^*_\mathit{S} (\rho^{AB}).
\ee
This bound can be easily interpreted. If the measured state has no entanglement, \ie  $r^*_\mathit{S} (\rho^{AB})=0$, then the probability distribution obtained by measuring it trivially has a separable realisation, so $\tilde{r}^{MDI}_\mathit{S}[p(a,b\vert \psi_x,\psi_y)]=0$. On the other hand, if Alice and Bob detect that $\tilde{r}^{MDI}_\mathit{S}[p(a,b\vert \psi_x,\psi_y)]> 0$, then they immediately conclude that the underlining state is entangled, and moreover can place a lower bound on the amount of entanglement, as measured by the robustness (with respect to $\mathit{S}$), that is necessary to explain the data.

\subsection{MDI lower bound on the negativity}

Another widely used entanglement measure of entanglement is the negativity \cite{Negativity}. Analogously to the device-independent estimation of negativity \cite{DInegativity} it is possible to put lower bound on the negativity in a measurement-device-independent way. The negativity $\mathcal{N}$ of some state $\rho^{AB}$ is defined as the sum of the absolute values of the non-positive eigenvalues of the partially transposed state $\rho^{T_A}$. It has been shown in \cite{Negativity} that it admits the following representation 
\begin{align}\label{negativity}
\mathcal{N}(\rho^{AB}) = \min_{\rho_+,\rho_-} &\quad \textrm{Tr}[\rho_-], \\ \nonumber
\textrm{s.t.} & \quad \rho^{AB} = \rho_+ - \rho_-,\\ \nonumber
& \quad \rho_{\pm}^{T_A} \geq 0.
\end{align}
Having in mind the decomposition $\rho^{AB} = \rho_+ - \rho_-$ it is possible to write the observed probabilities from the quantum inputs scenario in the following way
\begin{eqnarray*}
p(a,b\vert \psi_x,\psi_y) &=& \textrm{Tr}\left[\left(M_{a}^{A_0A}\tp M_b^{BB_0}\right)\left(\psi_x^{A_0}\tp \rho^{AB}\tp \psi_y^{B_0}\right)\right]\\
&=& q_+(a,b\vert \psi_x,\psi_y) - q_-(a,b\vert \psi_x,\psi_y), 
\end{eqnarray*}
where
\begin{eqnarray*}
q_+(a,b\vert \psi_x,\psi_y) &=& \textrm{Tr}\left[\left(M_{a}^{A_0A}\tp M_b^{BB_0}\right)\left(\psi_x^{A_0}\tp\rho_+\tp \psi_y^{B_0}\right)\right],\\
q_-(a,b\vert \psi_x,\psi_y) &=& \textrm{Tr}\left[\left(M_{a}^{A_0A}\tp M_b^{BB_0}\right)\left(\psi_x^{A_0}\tp\rho_-\tp \psi_y^{B_0}\right)\right].
\end{eqnarray*}
According to \eqref{negativity} the negativity can be obtained by minimizing $\tr[\rho_-]$, which can be written  as
\begin{align}
\sum_{a,b}&q_-(a,b\vert\psi_x,\psi_y) \nonumber \\
&= \textrm{Tr}\left[\left(\sum_a M_{a}^{A_0A}\tp \sum_b M_b^{BB_0}\right)\left(\psi_x^{A_0}\tp\rho_-\tp \psi_y^{B_0}\right)\right] \nonumber \\
&=\textrm{Tr}[\rho_-].
\end{align}
In order to estimate the negativity by an SDP optimization it is necessary to understand the form an effective POVM $\tilde{M}_{a,b}$ corresponding to a PPT state. We recall that for an arbitrary state an effective POVM must be positive and satisfy no-signalling principle, as encoded in \eqref{e:set M}, and for a separable state it also has to be a separable operator, as encoded in \eqref{e:set S}. An effective POVM corresponding to a PPT state, besides satisfying the no-signalling constraints, must also be a PPT operator. To see this consider partial transpose of an effective POVM 
\begin{eqnarray*}
\left(\tilde{M}_{a,b}^{A_0B_0}\right)^{T_{A_0}} = \textrm{Tr}_{AB}\left[\left(\left(M_{a}^{A_0A}\right)^{T_{A_0}}\tp M_b^{BB_0}\right)\left(\mathbb{I}^{A_0}\tp{\rho^{AB}}\tp \mathbb{I}^{B_0}\right)\right],\\ 
= \textrm{Tr}_{AB}\left[\left(\left(M_{a}^{A_0A}\right)^{T}\tp M_b^{BB_0}\right)\left(\mathbb{I}^{A_0}\tp\left(\rho^{AB}\right)^{T_A}\tp \mathbb{I}^{B_0}\right)\right].
\end{eqnarray*}
The second equality follows from the fact that $\Tr[A^TB] = \Tr[AB^T]$. Since the full transpose is a CPTP map, $\left(M_{a}^{A_0A}\right)^{T}$ is a positive operator and thus  $\left(\tilde{M}_{a,b}^{A_0B_0}\right)^{T_{A_0}}$ is positive if the state $\rho^{AB}$ is PPT, which is exactly the claim we wanted to prove. We will denote by $\mathcal{P}$ the set of effective POVMs that are also PPT, i.e.
\begin{multline}
\mathcal{P} =  \Big\{\{\tilde{M}_{a,b}^{A_0B_0}\}_{a,b}|\{\tilde{M}_{a,b}^{A_0B_0}\}_{a,b} \in \mathcal{M}, \left(\tilde{M}_{a,b}^{A_0B_0}\right)^{T_{A_0}} \geq 0\Big\}
\end{multline}
Now we have all the ingredients for the formulation of the SDP whose solution lower bounds the negativity of a state compatible with some observed probability distribution $p(a,b\vert \psi_x, \psi_y)$ in the quantum inputs scenario:
\begin{align}\label{SDPnegativity}
\mathcal{N}^{MDI}&[p(a,b\vert \psi_x,\psi_y)]= \min_{\{M_{\pm,a,b}^{A_0B_0}\}_{a,b}}\quad\sum_{a,b}q_-(a,b\vert\psi_x,\psi_y), \\ \nonumber
\text{s.t.}&\quad  p(a,b\vert\psi_x,\psi_y) = q_+(a,b\vert\psi_x,\psi_y)-q_-(a,b\vert\psi_x,\psi_y),\\ \nonumber
 &\quad q_{\pm}(a,b\vert\psi_x^{A_0},\psi_y^{B_0}) = \textrm{Tr}\left[M_{\pm,a,b}^{A_0B_0}\left(\psi_x\tp\psi_y\right)\right],\quad \forall a,b\\ \nonumber
&\quad\{M_{\pm,a,b}^{A_0B_0}\}_{a,b} \in \mathcal{P}
\end{align}\\

\section{Multipartite case}

In this section we will  generalize the previous entanglement detection and quantification techniques to the multipartite scenario. In Buscemi's paper \cite{Buscemi} there is an outline of the proof that all multipartite entangled states exhibit some kind of nonlocality when queried with quantum inputs. Moreover, the approach via entanglement witnesses is also explained in \cite{Branciard} (see also \cite{Zhao}), but as in the bipartite case the witness is tailored for a specific state. Here, as before, we are interested in the detection of multipartite entanglement without a priori knowledge of the system under study.\\ 

In the bipartite case we saw that the problem reduces to finding a separable effective POVM that returns the observed data when applied to the chosen set of inputs. This generalises to the multipartite case as follows: given a certain type of separability, we want to find the properties of the effective POVM which by acting on the given set of quantum inputs returns the observed probability distribution. As expected, we will show that the effective POVM should have the same type of separability properties as the underlying state. For the sake of simplicity we will consider the tripartite scenario, with the generalization to more parties straightforward.\\

The scenario involves three parties, Alice, Bob and Charlie, each of whom can input quantum systems in the states $\psi_x$, $\psi_y$ and $\psi_z$ respectively in their measuring devices, that subsequently provide classical outputs $a$, $b$ and $c$. The experiment is characterized by the set of joint probabilities of the form
\begin{multline}\label{eq:qimpprob}
p(a,b,c\vert \psi_x,\psi_y,\psi_z) = \textrm{Tr}\bigg[\big(M^{A_0A}_a\otimes M^{B_0B}_b \otimes M_{c}^{C_0C}\big)\\ 
\times \big(\rho^{ABC}\otimes \psi_x^{A_0}\otimes \psi_y^{B_0}\otimes \psi_z^{C_0}\big)\bigg],
\end{multline}
where $M^{A_0A}_a$ is a POVM Alice applies to the input $\psi_x$ and her share of the state $\rho^{ABC}$, and analogous for $M^{B_0B}_b$ and $M^{C_0C}_c$. In the same way as in the bipartite scenario it is useful to define an effective POVM
\be\label{eq:mtpeffectivePOVM}
\tilde{M}_{a,b,c}^{A_0B_0C_0} = \textrm{Tr}_{ABC}\bigg[\big(M^{A_0A}_a\otimes M^{B_0B}_b \otimes M_{c}^{C_0C}\big)\big(\rho^{ABC}\otimes \mathbb{I}^{A_0B_0C_0}\big)\bigg]
\ee
which allows one to write
\begin{multline}\label{eq:mtpeffectiveprob}
p(a,b,c\vert \psi_x,\psi_y,\psi_z) =\textrm{Tr}\big[ \tilde{M}_{a,b,c}^{A_0B_0C_0}(\psi_x^{A_0}\otimes \psi_y^{B_0}\otimes \psi_z^{C_0}) \big].
\end{multline}
As in the bipartite case, the effective POVM elements satisfy a number of constraints by construction, which play the role of no-signalling conditions. For example
\begin{equation}
\sum_a \tilde{M}_{a,b,c}^{A_0B_0C_0} = \mathbb{I}^{A_0} \otimes \tilde{M}_{b,c}^{B_0C_0},
\end{equation}
with $\tilde{M}_{b,c}^{B_0C_0}$ another effective POVM for Bob and Charlie. We will denote the set of effective POVMs which satisfy all such conditions in the tripartite case by $\mathcal{M}^{ABC}$.  

In what follows we will show that the entanglement properties of the effective POVM elements (\ref{eq:mtpeffectivePOVM}) are the same as the entanglement properties of the shared state $\rho^{ABC}$ .

Fully separable states can be written in the form $\rho^{ABC} = \sum_{\lambda}p_{\lambda}\rho_{\lambda}^A\tp \rho_{\lambda}^B \tp \rho_{\lambda}^C$ and if Alice, Bob and Charlie share such a state the corresponding effective POVM elements (\ref{eq:mtpeffectivePOVM}) will also be fully separable operators
\be\label{eq:fullsepeffPOVM}
\tilde{M}_{a,b,c}^{A_0B_0C_0} = \sum_{\lambda}p_\lambda M^{A_0}_{a|\lambda}\tp M^{B_0}_{b|\lambda} \tp M_{c|\lambda}^{C_0}
\ee
where $M_{a|\lambda}^{A_0} = \textrm{Tr}_A\left[M_a^{A_0A}\left(\mathbb{I}^{A_0}\tp\rho_{\lambda}^A\right)\right]$ is an effective POVM, and analogously for $M_{b|\lambda}^{B_0}$ and $M_{c|\lambda}^{C_0}$. Analogously to the bipartite case, we define a subset  $\mathcal{S}^{A|B|C}$ of all effective tripartite POVMs $\mathcal{M}^{ABC}$, which are also fully separable, $\mathcal{S}^{A|B|C} = \Big\{\{\tilde{M}_{a,b,c}^{A_0B_0C_0}\}_{a,b,c}|\{\tilde{M}_{a,b,c}^{A_0B_0C_0}\}_{a,b,c} \in \mathcal{M}^{ABC}, \tilde{M}_{a,b,c}^{A_0B_0C_0}= \sum_{\lambda} \tau_{a|\lambda}\otimes \chi_{b|\lambda}\otimes \omega_{c|\lambda},
\tau_{a|\lambda} \geq 0, \chi_{b|\lambda} \geq 0, \omega_{c|\lambda} \geq 0\Big\}$. With this in place, full separability of the shared state can thus be cast by the following feasibility problem:
\begin{align}
\textrm{given}& \quad \{p(a,b,c\vert \psi_x,\psi_y, \psi_z)\}_{a,b,c,x,y,z}, \nonumber \\
\textrm{find}& \quad \{\tilde{M}_{a,b,c}^{A_0B_0C_0}\}_{a,b,c} \label{eq:sdpFullSep} \\
\textrm{s.t.}&\quad p(a,b,c\vert \psi_x,\psi_y,\psi_z) = \textrm{Tr}\big[\tilde{M}_{a,b,c}^{A_0B_0C_0}(\psi_x^{A_0}\tp\psi_y^{B_0}\tp\psi_z^{C_0})\big] \nonumber\\& \qquad\qquad  \forall a,b,c,x,y,z \nonumber \\
&\quad \{\tilde{M}_{a,b,c}^{A_0B_0C_0}\}_{a,b,c}  \in \mathcal{S}^{A|B|C}\nonumber
\end{align}
In the similar way as in the bipartite scenario, it is necessary to choose an appropriate relaxation of the set of fully separable tripartite operators in order to turn this feasibility problem into an SDP. The set of operators which are PPT across all biparititons is one choice. A second option is to use the multipartite generalisation of the $k$-shareability hierarchy of SDPs \cite{DohertyReview}.\\

Tripartite states have a richer entanglement structure than bipartite states, such that even if the problem \eqref{eq:sdpFullSep} confirms that there is some entanglement in the system, a full entanglement characterization is not yet complete. It can happen, for example, that the entanglement is shared only between  two parties. States which have such entanglement structure are called separable across a certain bipartition. For example the state $\rho^{ABC} = \sum_{\lambda}p_{\lambda}\rho_{\lambda}^{AB}\tp \rho_{\lambda}^C$ is separable with respect to the bipartition $AB\vert C$. A state is biseparable if it can be written as a convex combination of states that are separable with respect to different bipartitions:
\begin{multline}\label{eq:bspdecomp}
\rho^{ABC} = \sum_{\lambda}p_{\lambda}^{A|BC}\rho_{\lambda}^A \tp \rho_{\lambda}^{BC} \\+ \sum_{\mu}p_{\mu}^{B|AC}\rho_{\mu}^{B} \tp \rho_{\mu}^{AC} + \sum_{\nu}p_{\nu}^{C|AB}\rho_{\nu}^{AB} \tp \rho_{\nu}^{C}.
\end{multline}
The strongest form of entanglement that can be present in a tripartite system is genuine multipartite entanglement (GME). A state $\rho^{ABC}$ is genuinely multipartite entangled if it is not biseparable.

Let us assume that the state shared between Alice, Bob and Charlie is biseparable (\ref{eq:bspdecomp}). In that case the effective POVM reads
\begin{multline}\label{eq:effPOVMgme}
\tilde{M}_{a,b,c}^{A_0B_0C_0} = \sum_{\lambda}p_{\lambda}^{A|BC}M_{a|\lambda}^{A_0}\tp M_{b,c|\lambda}^{B_0C_0} \\
+  \sum_{\mu}p_{\mu}^{B|AC}M_{b|\mu}^{B_0}\tp M_{a,c|\mu}^{A_0C_0} +  \sum_{\nu}p_{\nu}^{AB|C}M_{a,b|\nu}^{A_0B_0}\tp M_{c|\nu}^{C_0},
\end{multline}
where
\begin{eqnarray*}
M_{a|\lambda}^{A_0} &=& \textrm{Tr}_A\left[M_a^{A_0A}\left(\psi_x^{A_0}\tp \rho_{\lambda}^{A}\right)\right],\\ 
M_{b,c|\lambda}^{B_0C_0} &=& \textrm{Tr}_{BC}\left[\left(M_b^{B_0B}\tp M_c^{CC_0}\right)\left(\psi_y^{B_0}\tp \rho_{\lambda}^{BC}\tp\psi_z^{C_0}\right)\right]
\end{eqnarray*}
and analogously for all other operators. Thus, the fact that the state $\rho^{ABC}$ is biseparable implies that the operators \eqref{eq:effPOVMgme} are also biseparable. 

With this structure in mind it is possible to construct a feasibility problem to test whether an observed probability distribution in the quantum input scenario can be obtained with a biseparable state:
\begin{align}
\textrm{given}& \quad \{p(a,b,c\vert \psi_x,\psi_y, \psi_z)\}_{a,b,c,x,y,z}, \nonumber \\
\textrm{find}& \quad \{\tilde{M}_{a|b,c}^{A_0B_0C_0}, \tilde{M}_{b \vert a,c}^{A_0B_0C_0}, \tilde{M}_{c \vert a,b}^{A_0B_0C_0}\}_{a,b,c} \label{eq:sdp} \\
\textrm{s.t.}&\quad p(a,b,c\vert \psi_x,\psi_y,\psi_z) = \textrm{Tr}\big[\tilde{M}_{a,b,c}^{A_0B_0C_0}(\psi_x^{A_0}\tp\psi_y^{B_0}\tp\psi_z^{C_0})\big] \nonumber\\
& \qquad\qquad  \forall a,b,c,x,y,z \nonumber \\
&\quad \tilde{M}_{a,b,c}^{A_0B_0C_0} = \tilde{M}_{a\vert b,c}^{A_0B_0C_0} + \tilde{M}_{b \vert a,c}^{A_0B_0C_0} + \tilde{M}_{c \vert a,b}^{A_0B_0C_0}\quad \forall a,b,c  \nonumber \\
&\quad \{\tilde{M}_{a\vert b,c}^{A_0B_0C_0}\}_{a,b,c} \in \mathcal{S}^{A\vert B,C}, \quad \{\tilde{M}_{b\vert a,c}^{A_0B_0C_0}\}_{a,b,c} \in \mathcal{S}^{B\vert A,C},\nonumber \\
&\quad \{\tilde{M}_{c\vert a,b}^{A_0B_0C_0}\}_{a,b,c} \in \mathcal{S}^{C\vert A,B} .\nonumber
\end{align}
where $\mathcal{S}^{A\vert B,C}$ denotes the subset of effective tripartite POVMs $\mathcal{M}^{ABC}$ that are also separable across the bipartition $A\vert B,C$ and analogously for $\mathcal{S}^{B\vert A,C}$  and  $\mathcal{S}^{C\vert A,B}$. Once more, by replacing the sets $\mathcal{S}$ by the set of PPT operators or operators having $k$-symmetric extension the above problem becomes an instance of a SDP. \\

Quantification of multipartite entanglement can be performed in a similar manner as in the bipartite scenario. Namely, one can lower bound the robustness of genuine multipartite entanglement, or simply robustness of multipartite entanglement, by defining MDI multipartite nonlocality robustness analogously to \eqref{mdirob}.

\section{Randomness from quantum inputs}

Nonlocal correlations, as proven by Bell's theorem \cite{Bell}, cannot be explained by any classical, deterministic model or a convex combinations of such models.  Consequently, a violation of a Bell inequality can be used to certify that the data generated is intrinsically random. This reasoning led to the development  of the protocols for so-called device-independent randomness certification \cite{DI2,DI3,DI4,DI5}. In these protocols the amount of (global) randomness stemming from some Bell experiment is characterized by the guessing probability $G_{x,y}$ with which an external eavesdropper can guess a pair of outcomes observed by Alice and Bob when they make measurements $x$ and $y$. A lower bound on the guessing probability is
\be\label{eq:Gprob}
G_{x^*,y^*} = \max_{a,b}p(a,b\vert x^*,y^*).
\ee
and this is the best that an external observer uncorrelated with Alice and Bob can guess. However, an eavesdropper, usually named Eve, can have side-information -- a system that is correlated (or even entangled) with the state of Alice and Bob. In principle she could have even provided all the measuring devices, and can possibly achieve much better guessing probability than the lower bound  \eqref{eq:Gprob}. Thus the aim of a device-independent randomness estimation protocol is to quantify the randomness of Alice's (and/or Bob's) measurement outcomes by optimizing over all possible eavesdropping strategies of Eve compatible with the obtained Bell inequality violation. The scenario assumes that by sharing a tripartite state, and performing some measurement on her share, Eve steers the state of Alice and Bob. Her strategy is to perform a measurement such that her outcome, denoted by $e$, will give her the highest probability to guess the pair of outcomes for one particular choice of measurements for Alice and Bob. In such a scenario calculating Eve's guessing probability,  if the obtained value of a Bell expression $\sum_{a,b,x,y}I_{a,b,x,y}p(a,b\vert x,y)$ is equal to $\beta$  can be cast as the following optimisation problem
\begin{align}\label{eq:randomnessSDP}
G_{x^*,y^*} = & \max_{\{p(a,b,e|x,y)\}_{a,b,e,x,y}}\sum_e p(a,b,e=(a,b)\vert x^*,y^*),\\
& \textrm{s.t} \quad \sum_{a,b,e,x,y}I_{a,b,x,y}p(a,b,e\vert x,y) = \beta; \nonumber \\
& \{p(a,b,e\vert x,y,z)\}_{a,b,e,x,y} \in \mathcal{Q}. \nonumber
\end{align}
where $\mathcal{Q}$ denotes the set of quantum behaviours, i.e. the set of all $\{p(a,b,e|x,y)\}_{a,b,e,x,y}$ that can arise by performing local measurements on a tripartite quantum state. This program gives the highest probability with which Eve's outcome $e$ is the same as Alice's and Bob's outcomes, $a$ and $b$, for some specific pair of inputs $x^*$ and $y^*$, with the constraints that the overall probabilities must be compatible with quantum mechanics and the observed violation of a Bell inequality. In general this problem cannot be solved exactly, due to the set $\mathcal{Q}$ having no known simple characterisation (in particular since it implicitly contains all behaviours compatible with any quantum state and measurements in a Hilbert space of any dimension) however by using the Navascues-Pironio-Acin (NPA) hierarchy of SDP relaxations of the quantum set of behaviours \cite{NPA}, computable upper bounds can be placed on the guessing probability.\\ 

One generalization of this protocol to the quantum-input scenario under the name measurement-device-independent randomness certification has been introduced in \cite{Banik}. Analogously to the ability to detect entanglement of all entangled states, even those that do not violate any Bell inequality, the authors of \cite{Banik} prove that it is possible to extract randomness from local entangled states in a measurement-device-independent way. They use the analogue of the program \eqref{eq:randomnessSDP} with the constraint that the probabilities $p(a,b\vert \psi_x,\psi_y)$ violate the inequality corresponding to a specific MDIEW. As already noted in the previous text, a MDIEW is usually constructed with respect to a specific entangled state. It can nevertheless be used to check if some other entangled state in principle can be useful for randomness extraction. However, as the source providing Alice's and Bob's shared state is uncharacterised it may not be clear which witness should be used, and therefore it is desirable to have a method to certify randomness that does not rely on a specific MDIEW. 
Another way to certify randomness in the quantum input scenario is the subject of \cite{Cao}. In this approach the source has to prepare a tomographically complete set of inputs which are used to perform quantum process tomography of the measurement device. Randomness is generated by measuring one of the prepared quantum inputs by the characterized POVM. A method do quantify the amount of randomness is presented for two-outcome POVMs. This approach was used to experimentally generate randomness in a measurement-device-independent manner \cite{ExpMDIRG}.\\
 
In the following we will show the way to quantify the amount of randomness resulting from an experiment with quantum inputs without assuming the underlying state. This can be seen as the generalisation of the approach of \cite{Olmo,Bancal} from the standard Bell scenario to the quantum inputs scenario. The protocol works for measurements with arbitrary number of outputs and the set of quantum inputs does not have to be tomographically complete, which makes it more general than \cite{Cao}. \\

Before presenting the more general approach to randomness estimation in the quantum inputs scenario, let us consider in more detail the essential novelty of this scenario, which is the fact that before guessing the measurement outcome Eve has to guess the input state. Due to this it is not only possible to extract randomness from local entangled states as observed in \cite{Banik}, but also from a single black-box, i.e. without the use of entanglement. \\

This leads us to the change of scenario: now we have only one party, Alice, who has a characterised device which prepares quantum input states $\ket{\psi_x}$. Alice measures these states using an uncharacterised black box, modelled by the POVM $\{M_a\}_a$ and obtains some outcomes $a$. By repeating the process she can calculate the set of probabilities $p(a\vert \psi_x)$, to get an outcome $a$ when the quantum input is $\ket{\psi_x}$. The question is how random Alice's outputs are for Eve. In some cases Alice can be sure that her outcomes are genuinely random. If the set of quantum inputs is tomographicaly complete Alice can perform process tomography and exactly learn which POVM her black-box is applying. In the special case when the obtained POVM is extremal Eve cannot be correlated with Alice's experiment because extremal POVMs cannot be decomposed as a convex combination of other POVMs. Therefore Eve's guessing probability is obtained simply from \eqref{eq:Gprob} (restricted to a single party Alice, instead of Alice and Bob). An extremal $d$-dimensional POVMs cannot have more than $d^2$ outcomes \cite{dariano}, which means that when preparing qubit quantum inputs Alice at best can get $2$ bits of randomness. One example is the case when Alice prepares the following informationally complete set of quantum inputs $\{\frac{\mathbb{I}}{2},\ket{0},\frac{\ket{0}+\ket{1}}{\sqrt{2}},\frac{\ket{0}+i\ket{1}}{\sqrt{2}}\}$ and observes the probability distribution which corresponds to measuring these inputs with an extremal four-outcome tetrahedral POVM. The probabilities to get any of the four outcomes when measuring the input which is in the maximally mixed state $\frac{\mathbb{I}}{2}$ is equal to $0.25$, which corresponds to $2$ bits of randomness.\\

In the general case, when the set of quantum inputs is not tomographically complete, or the applied POVM is not extremal it is still possible to construct an SDP-based estimation of randomness analogous to \eqref{eq:randomnessSDP}. Since the measuring device is uncharacterised it has to be assumed that it was possibly prepared by Eve. In that case Eve can be quantumly correlated with the box. That is, she can prepare an ancillary entangled pair of particles $\rho^{AE}$, and place half inside the box, while keeping the other half. In each round Alice's box then performs a joint measurement $M_{a}^{A_0A}$, where $a$ corresponds to Alice's outcome, and Eve performs a measurement $N_e^E$, with outcome $e$. Thus, the probability for Alice to get outcome $a$ and Eve to get $e$, when the quantum input is $\ket{\psi_x}^{A_0}$ is
\ba\label{eq:SingleBoxProbability}
p(a,e\vert \psi_x) &=& \textrm{Tr}\left[(M_{a}^{A_0A}\tp N_e^E)\left(\psi_x^{A_0}\tp \rho^{AE}\right)\right] \\ \nonumber
&=&\tr[\tilde{M}_{a,e}^{A_0}\psi_x^A],
\ea
where $\tilde{M}_{a,e}$ is an effective POVM which satisfies the relation
\be\label{eq:SingleBoxEffPovm}
\tilde{M}_{a,e}^{A_0} = \textrm{Tr}_{AE}\left[(M_{a}^{A_0A}\tp N_e^E)\left(\mathbb{I}^{A_0}\tp \rho^{AE}\right)\right].
\ee
From this relation, it follows that the effective POVM satisfies the relation
\begin{equation}
	\sum_a \tilde{M}_{a,e}^{A_0} = p(e)\mathbb{I}^{A_0},
\end{equation}
where $p(e) \geq 0$. As in the above, this can be seen as the no-signalling constraint from Alice to Eve.

With the above in place, the optimisation problem which bounds the guessing probability of Eve is
\begin{align}\label{eq:SingleBoxguessingProb}
G_{x^*} = \max_{\{\tilde{M}_{a,e}^{A_0}\}_{a,e}} &\tr\sum_e \tilde{M}_{a=e,e}^{A_0}\psi_{x^*}^{A_0},\\
 \textrm{s.t}& \quad \tr\sum_{e}\tilde{M}_{a,e}^{A_0}\psi_x^{A_0} = p(a\vert \psi_x), \quad \forall x,a;\nonumber \\
&\quad\sum_a \tilde{M}_{a,e}^{A_0} = p(e)\mathbb{I}^{A_0}, \quad \sum_ep(e) = 1, \quad \forall e. \nonumber
\end{align}
The objective function maximizes the probability for Eve to guess the outcome $a$ when Alice measures the quantum input $\psi_x^{A_0}$, by optimizing over all possible effective POVMs $\tilde{M}_{a,e}^{A_0}$. The first constraint ensures that that effective POVMs are in accordance with the observed probability distribution, while the second imposes no-signalling and  completeness of the measurement. For any probability distribution obtained by measuring some set of quantum inputs, the above optimisation problem, which is an SDP, gives the guessing probability, and thus the randomness of the outcomes.\\

A similar analysis can be applied in the bipartite case. As mentioned above, in the standard Bell scenario one does not need to consider a specific Bell inequality in order for randomness estimation, but rather can use the full nonlocal behaviour (set of correlations) obtained in a Bell experiment \cite{Olmo,Bancal}. In what follows we will generalize this method, providing an alternative way to quantify randomness in a measurement-device-independent manner. Specifically we will show that it is possible to certify randomness even when Alice and Bob share a separable state.\\

In the scenario with quantum inputs Eve again distributes a state to Alice and Bob with which she is entangled. By performing a local POVM $N_e$ on her share, and conditioned on the outcome $e$, she prepares an subnormalized state $\rho_e^{AB}$, whose norm is equal to the probability for Eve to obtain the outcome, $\textrm{Tr}\rho_e^{AB} = p(e)$. The full joint probability is given by
\ba\nonumber
p(a,b,e\vert \psi_x,\psi_y) &=& \textrm{Tr}\left[\left(M_a^{A_0A} \tp M_b^{BB_0}\right)\left(\psi_x^{A_0}\tp \rho_e^{AB}\tp \psi_y^{B_0}\right)\right]\\ \label{eq:probEve}
&=& \textrm{Tr}\left[\tilde{M}_{a,b,e}^{A_0B_0} \left(\psi_x^{A_0}\tp \psi_y^{B_0}\right)\right]
\ea
where $\tilde{M}_{a,b,e}$ is an effective POVM defined by
\be\label{eq:effectivePOVMEve}
\tilde{M}_{a,b,e}^{A_0B_0} = \textrm{Tr}_{AB}\left[\left(M_{a}^{A_0A}\tp M_b^{BB_0}\right)\left(\mathbb{I}^{A_0}\tp \rho_e^{AB}\tp \mathbb{I}^{B_0}\right)\right].
\ee
These effective POVMs, apart from satisfying the completeness relation $\sum_{a,b,e}\tilde{M}_{a,b,e}^{A_0B_0} = \mathbb{I}^{A_0B_0}$ also satisfy the no-signalling constraints
\begin{equation}
	\begin{split}
		\sum_a \tilde{M}_{a,b,e}^{A_0B_0} = \mathbb{I}^{A_0} \otimes \tilde{M}_{b,e}^{B_0}, \\
		\sum_b \tilde{M}_{a,b,e}^{A_0B_0} = \tilde{M}_{a,e}^{A_0} \otimes \mathbb{I}^{B_0}, 
	\end{split}
\end{equation}
where $\{\tilde{M}_{b,e}^{B_0}\}_b$ and $\{\tilde{M}_{a,e}^{A_0}\}_a$ are sub-normalised POVMs (for Bob and Alice respectively), for all values of $e$, with the same normalisation for each $e$, i.e. $\sum_b \tilde{M}_{b,e}^{B_0} = p(e)\mathbb{I}^{B_0}$ and $\sum_a \tilde{M}_{a,e}^{A_0} = p(e)\mathbb{I}^{A_0}$.

Like in the standard Bell scenario Eve's optimal strategy is to perform a measurement such that the outcome $e$ will be equal to the pair $(a,b)$ with as high a probability as possible, for some specific pair of quantum inputs $\psi_{x^*}^{A_0}$ and $\psi_{y^*}^{B_0}$. Eve's optimal guessing probability is then the solution of the following SDP
\begin{align}\label{eq:QIguessingProb}
G_{x^*,y^*}^{MDI}=\max_{\{\tilde{M}_{a,b,e}\}_{a,b,e}^{A_0B_0}} \quad & \tr\sum_e\tilde{M}_{a,b,e=(a,b)}^{A_0B_0}\left(\psi_{x^*}^{A_0}\tp\psi_{y^*}^{B_0}\right),\\
\textrm{s.t}\quad  & \tr\sum_{e}\tilde{M}_{a,b,e}^{A_0B_0}\left(\psi_x^{A_0}\tp \psi_y^{B_0}\right) \nonumber \\ 
&\qquad \qquad = p(a,b\vert \psi_x,\psi_y), \forall x,y,a,b,\nonumber \\
& \tilde{M}_{a,b,e}^{A_0B_0} \geq 0 \qquad \forall a,b,e, \nonumber \\
& \sum_a \tilde{M}_{a,b,e}^{A_0B_0} = \mathbb{I}^{A_0}\tp \tilde{M}_{b,e}^{B_0} \quad \forall b,e,\nonumber \\
&\sum_b \tilde{M}_{a,b,e}^{A_0B_0} = \tilde{M}_{a,e}^{A_0} \tp \mathbb{I}^{B_0} \quad \forall a,e; \nonumber \\
&\sum_b \tilde{M}_{b,e}^{B_0} = p(e)\mathbb{I}^{B_0}\quad \forall e, \nonumber \\
&\sum_a \tilde{M}_{a,e}^{A_0} = p(e)\mathbb{I}^{A_0}\quad \forall e, \nonumber \\  
&\sum_ep(e) = 1. \nonumber
\end{align}
The objective function is the total probability for Eve to guess Alice's and Bob's outputs for some specific pair of quantum inputs $\psi_{x^*}^{A_0}$ and $\psi_{y^*}^{B_0}$. The first constraint imposes consistency with the observed behaviour in the experiment, while the remaining constraints ensure a valid effective POVM (which is normalised and no-signalling).\\

\begin{figure}
\centering
\includegraphics[width=0.9\columnwidth]{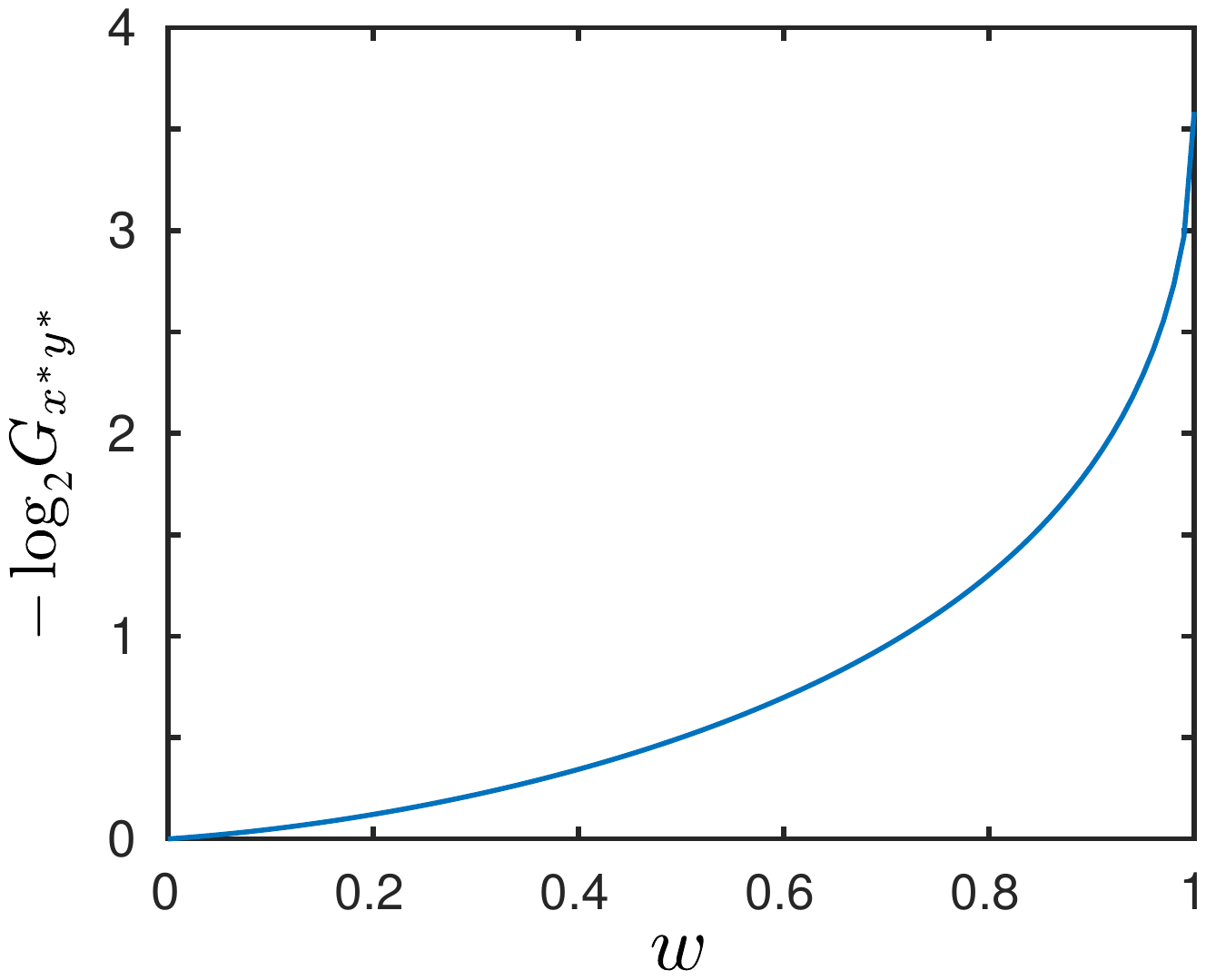}
\caption{\label{f:global randomness} Min-entropy ($-\log G_{x^*y^*}$) versus noise $w$ for the probability distribution that arises by performing Bell state measurements on the two-qubit Werner state, and quantum inputs along the vertices of a tetrahedron on the Bloch sphere. For all $w\neq 0$ (i.e. whenever the state is not equal to the maximally mixed state), then randomness can be certified. This graph was produced using the \textsc{cvx} \cite{cvx} package for Matlab, and the toolbox \textsc{qetlab} \cite{qetlab}. }
\end{figure}

As an example this program can be used to obtain the optimal guessing probability compatible with the probability distribution which arises by Alice and Bob performing Bell state measurements on a shared Werner state $\ket{\Psi}\bra{\Psi} = w\ket{\Phi^+}\bra{\Phi^+} + (1-w)\frac{\mathbb{I}}{4}$ and with quantum inputs corresponding to the vertices of tetrahedron on the Bloch sphere. The resulting guessing probability in terms of parameter $w$ is presented in Fig.~\ref{f:global randomness}.

 The correlations obtained on the maximally entangled state ($w = 1$) allow to extract four bits of randomness, because the guessing probability is $\frac{1}{16}$. As in \cite{Banik} some randomness can be observed from all entangled Werner states, even those admitting a local model. What may be particularly surprising is that actually all Werner states except for the maximally mixed state manifest some randomness. As commented earlier, intuitively this can be explained by the fact that Eve cannot with certainty guess which quantum input was used, which makes the  probability distribution random even when there is no nonlocality at all.

\section{Discussion}

In this work we have provided new insights into entanglement and randomness detection and quantification in the measurement-device-independent scenario. As explained, this scenario differs from the well-known device-independent scenario by the fact that the parties possess a characterised device that can prepare quantum system in some defined quantum states. The scenario in which some parties do not trust their sources and measurements but have a characterised preparation device is not so uncommon in quantum information processing and has been used for constructing protocols for universal blind quantum computation \cite{bqc}. In particular, we showed how one can estimate the values of two widely used entanglement measures, robustness-based quantifiers of entanglement, and entanglement negativity. Furthermore we showed how entanglement detection and quantification can be performed in quantum networks (i.e. in situations involving multiple parties, not just two).\\

On the other hand we showed how possessing a characterised preparation device can decrease adversarial power in guessing measurement outputs. Already a single party which can prepare specific states and measure them with a black box can extract two bits of randomness. Two parties sharing some quantum state can extract randomness even when they do not share any entanglement. \\

There are a number of interesting directions for future work. First is to study whether the results presented here for the measurement-device-independent scenario can be adapted to other device-independent scenarios. A second interesting avenue is to explore the prospects with regard to full quantum state recovery. A third direction is to focus on mixed quantum inputs -- in which can one can imagine that the Eavesdropper holds a purification. It is interesting to ask how this affects entanglement detection, and randomness estimation.

\textit{Note added}: After finishing this work we learnt about independent work of Rosset et al. \cite{Rosset2} also dealing with measurement-device-independent entanglement quantification.

\section{acknowledgements}
This work was supported by the Ram\'on y Cajal fellowship, Spanish MINECO (QIBEQI FIS2016-80773-P and Severo Ochoa SEV-2015-0522), the AXA Chair in Quantum Information Science, Generalitat de Catalunya (SGR875 and CERCA Programme), Fundaci\'{o} Privada Cellex and ERC CoG QITBOX, and a Royal Society University Research Fellowship.

\end{document}